\documentclass[12pt]{elsart}
\usepackage{epsfig,xspace}
\newcommand{\mum  }{\ensuremath{{\mu m}}\xspace}
\newcommand{\fc   }{\ensuremath{\mathrm{fC}}\xspace}
\newcommand{\mvfc }{\ensuremath{\mathrm{mV/fC}}\xspace}
\newcommand{\eenc }{\ensuremath{\mathrm{e^-ENC}}\xspace}
\newcommand{\gev  }{\ensuremath{\mathrm{GeV}}\xspace}
\makeatletter
\def\ps@copyright{}
\makeatother
\begin{document}
{\bf \hfill MPP-2003-120}
\begin{frontmatter}
\title{End-cap Modules for the ATLAS SCT\thanksref{label1}}
\thanks[label1]{Invited talk presented at the RD03 conference in Florence, 
                29 September 2003.}
\author{Richard Nisius\thanksref{label2}}
\thanks[label2]{E-mail address: nisius@mppmu.mpg.de}
\address{Max-Planck-Institut f\"ur Physik (Werner-Heisenberg-Institut), 
         F\"ohringer Ring 6, D-80805 M\"unchen, Germany.}
\address{Representing the ATLAS SCT Collaboration}
%
%------------------------------------------------------------------------------
%
\begin{abstract}
 The performance of prototype end-cap modules of the ATLAS SemiConductor 
 Tracker (SCT) are discussed. The results are obtained in stand-alone as 
 well as test beam measurements performed on modules both before and after
 irradiation with protons with a total dose corresponding to the expectation 
 for ten years of operation at the LHC.
  Finally, the present status of construction is summarised.
\end{abstract}
%
%------------------------------------------------------------------------------
%
\end{frontmatter}
%
%------------------------------------------------------------------------------
%
\section{Introduction}
\label{sec:intro}
\begin{figure}[htb]
\begin{center}
{\includegraphics[width=0.49\linewidth,clip]{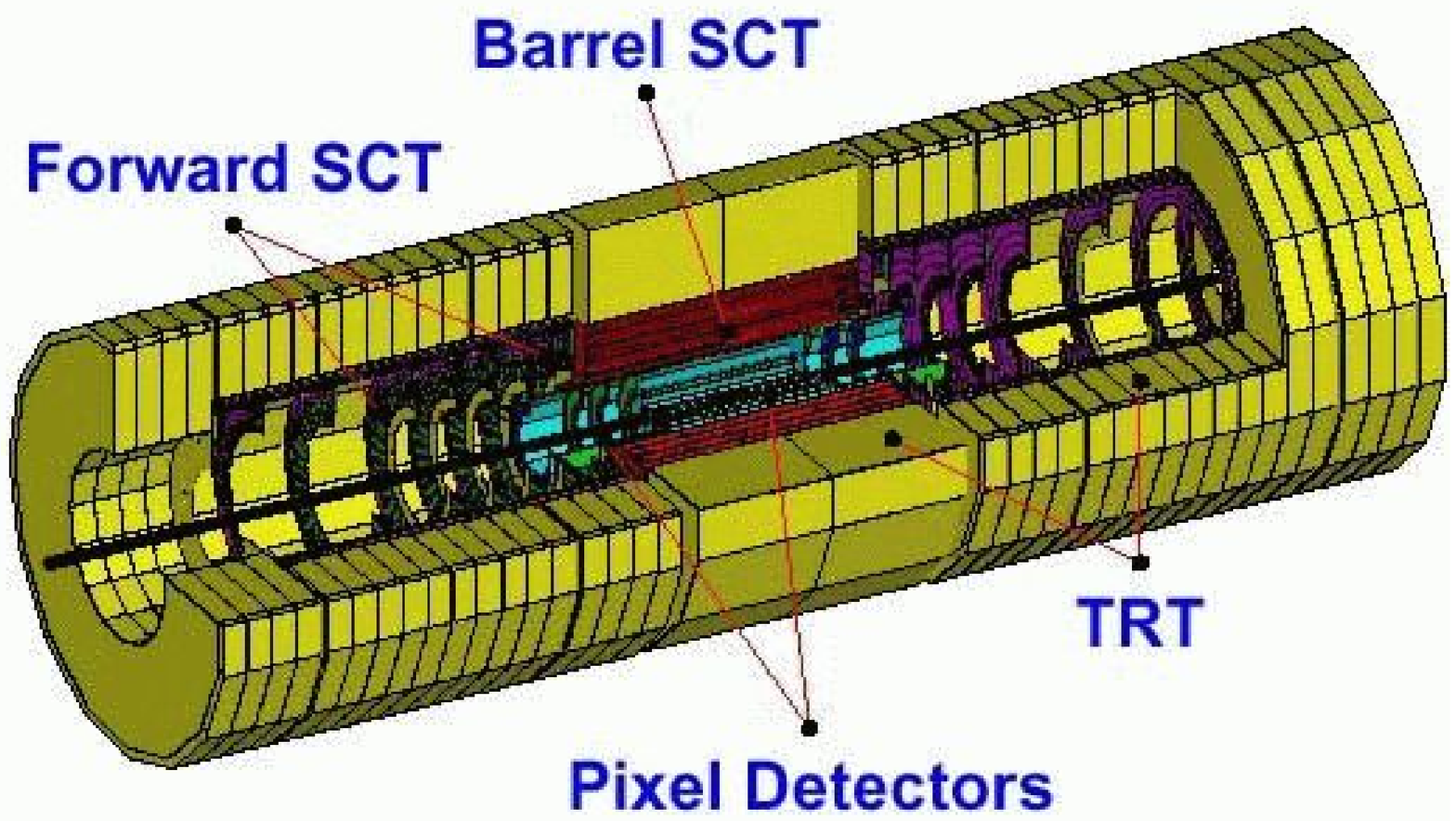}}
{\includegraphics[width=0.49\linewidth,clip]{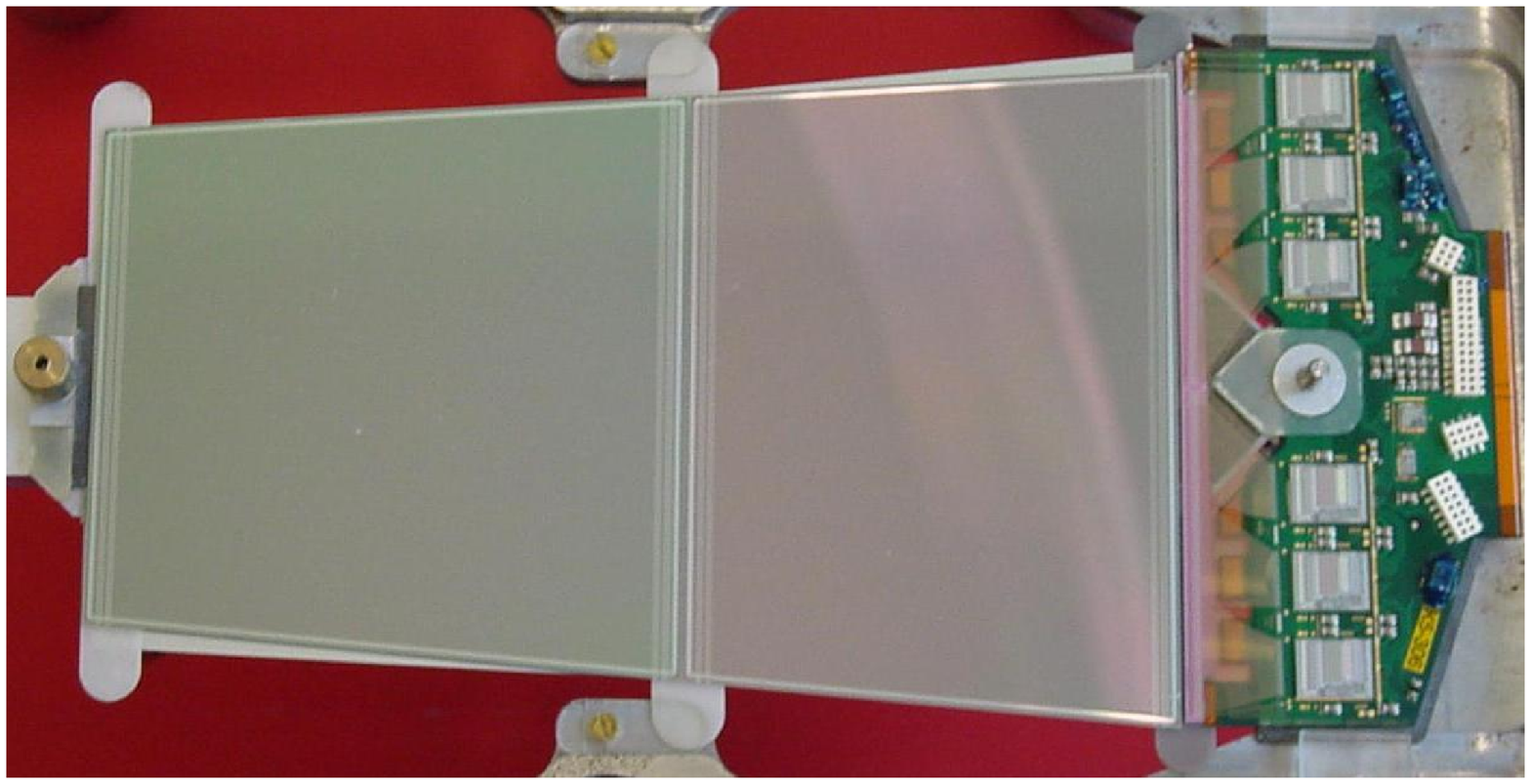}}
\begin{picture}(0,0)
\put(-390,120){\large\bf a)}
\put(-190,120){\large\bf b)}
\end{picture}
\caption{{\it The general layout of the inner detector a), and a middle 
              module for the end-cap b).}
        }\label{fig01}
\end{center}
\end{figure}

 The ATLAS~\cite{ATLAS} experiment at the Large Hadron Collider LHC 
 will contain a large silicon strip detector as part of the inner
 tracking detector.
 This SemiConductor Tracker (SCT)~\cite{ID} consists of a barrel
 and two forward end-caps, see Fig.~\ref{fig01}a. 
 The barrel and end-cap parts are equipped with about 2000 modules each.
%
%------------------------------------------------------------------------------
%------------------------------------------------------------------------------
%
 The modules are made from single-sided p-on-n silicon strip 
 sensors~\cite{SENSORS} glued back-to-back, with a 40~mrad stereo angle, on a 
 graphite support structure, and read-out in binary mode by custom made ASICs.
 For the end-cap modules the 12 ASICs per module, together with components for 
 the optical transmission of commands and data, are mounted on a double-sided
 hybrid produced from a 6 layer copper/Kapton flex.
 Three different end-cap module types (inner, middle and outer), 
 see Fig.~\ref{fig01}b for an example, made from five different wedge 
 shaped sensor designs are needed.
 The sensors are 285~\mum thick and contain 768 read-out strips with
 a pitch varying in the range from 55 to 95~\mum.
 The inner modules only contain two sensors, whereas the middle and outer
 modules each have four sensors of slightly different shape, such that all 
 module types have different sensor capacitance.
 Within the tight constraints imposed by the need for radiation hardness, 
 high rate capability, low mass, low cost and overlap between neighbouring 
 modules the design~\cite{ECMOD} has been optimised for a thermal split 
 between the read-out ASICs and the sensors.
%
%------------------------------------------------------------------------------
%
\section{Performance of prototype end-cap modules}
\label{sec:resul}
 A number of prototype end-cap modules has been built to verify the design
 concerning the thermal, mechanical and electrical performance. 
 Some recent results from these modules are reported here. 
 Earlier results can be found in~\cite{ECMOD}, and the performance in
 multi module tests, together with the module behaviour under 
 deliberate injection of external noise, is discussed in~\cite{FERRARI}.
 \par
%
%------------------------------------------------------------------------------
\begin{figure}[htb]
\begin{center}
{\includegraphics[width=1.00\linewidth,clip]{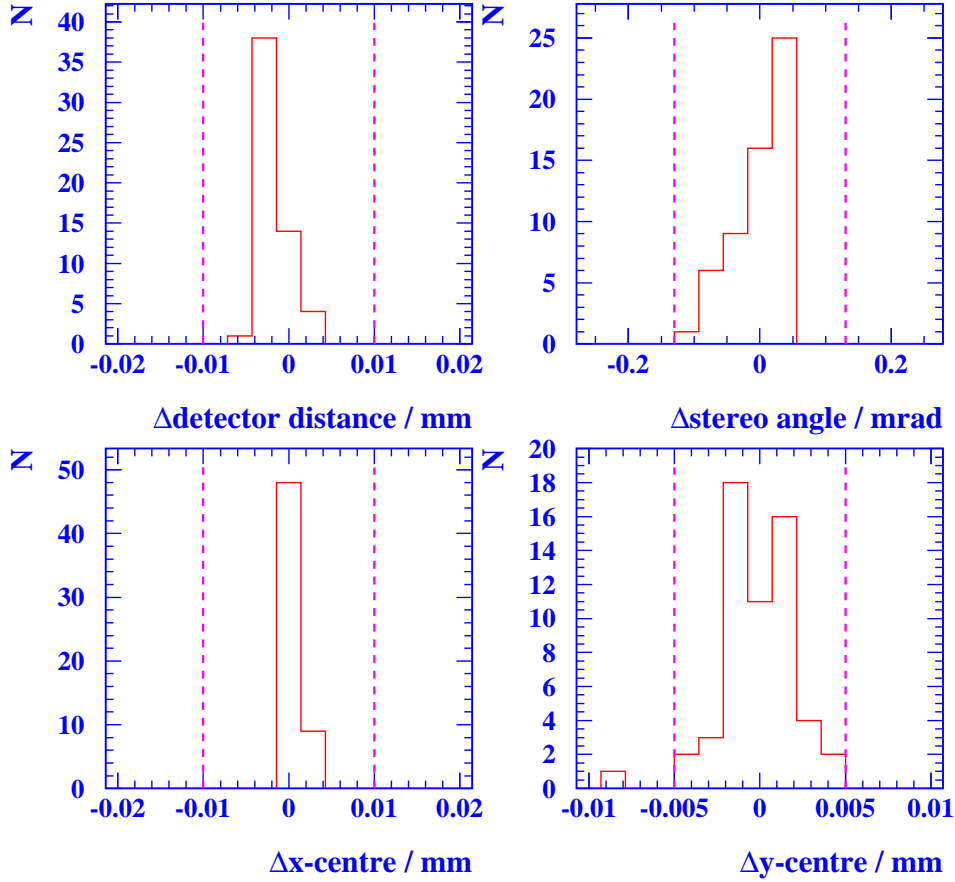}}
\caption{{\it Some mechanical parameters for prototype end-cap modules.}
        }\label{fig02}
\end{center}
\end{figure}
%------------------------------------------------------------------------------
%
 To achieve a precise tracking the position and orientation of the individual 
 sensors has to be known with high accuracy.
 To facilitate the detector calibration with tracks, and to be able to
 reliably contribute to the fast trigger, the module design 
 aims for 'equal' modules such that initially only modules and not 
 individual sensors have to be aligned~\cite{ID}. 
 This imposes strict requirements on the module construction, the strongest 
 being an only $\pm 5$~\mum tolerance on the reproducibility of the positioning
 of the sensors perpendicular to the strips.
 Using precision stages for the sensor alignment and optimising
 the various steps in the module construction these requirements can be met
 with high yield. This is demonstrated for some geometrical parameters
 in Fig.~\ref{fig02}.
 The distributions of deviations from the nominal values shown are well 
 within the tolerances, which are indicated by the dashed lines.
 \par
%
%------------------------------------------------------------------------------
\begin{figure}[htb]
\begin{center}
{\includegraphics[width=0.49\linewidth,clip]{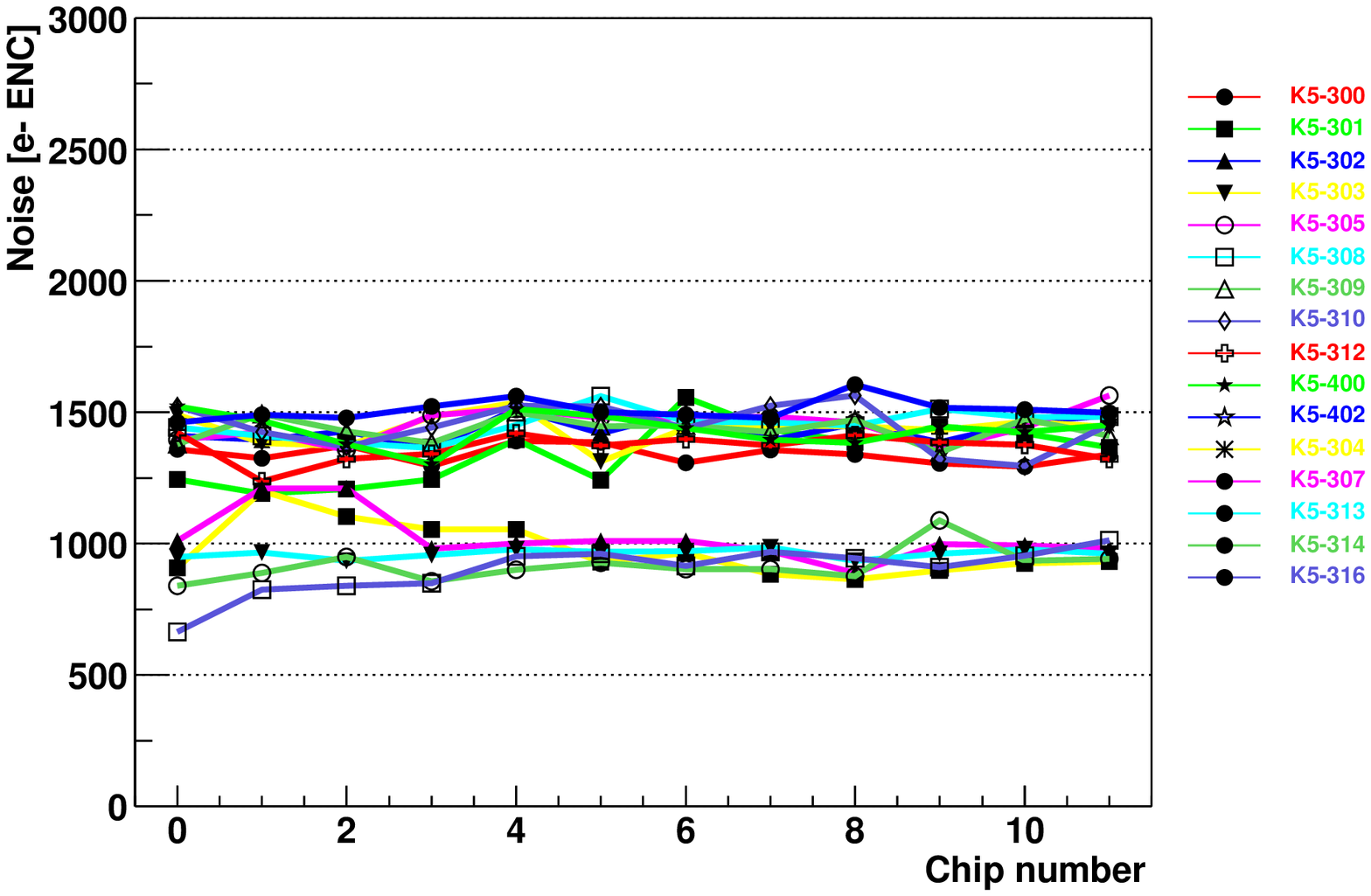}}
{\includegraphics[width=0.49\linewidth,clip]{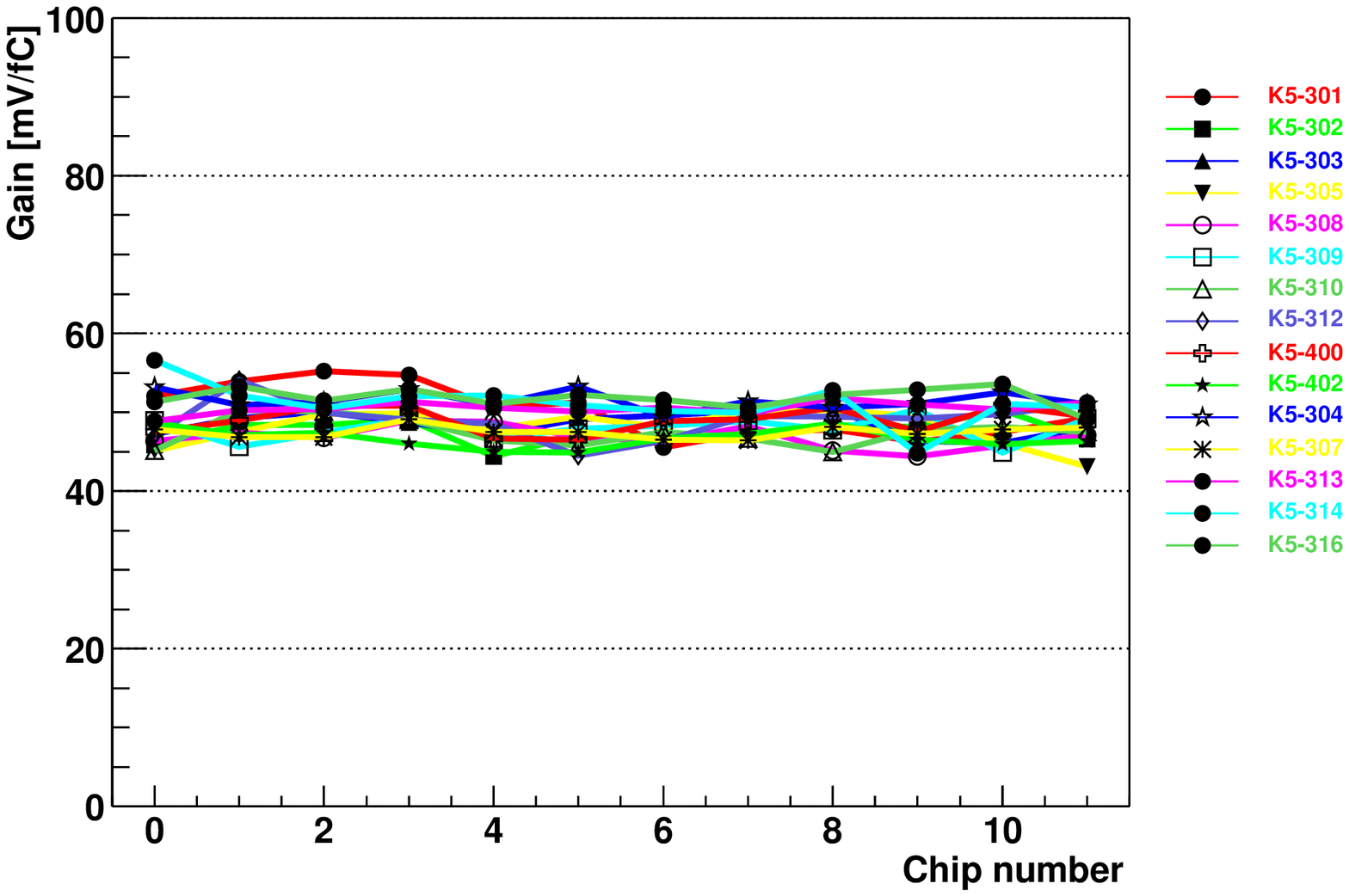}}
\begin{picture}(0,0)
\put(-355,127){\large\bf a)}
\put(-160,127){\large\bf b)}
\end{picture}
\caption{{\it The average noise a) and gain b) per chip 
              for prototype end-cap modules.}
        }\label{fig03}
\end{center}
\end{figure}
%------------------------------------------------------------------------------
%
 A key issue for the binary read-out scheme is the stability of the modules
 with respect to noise and gain.
 The chip-to-chip and module-to-module variation of noise and gain is 
 small, Fig.~\ref{fig03}.
 Since the inner modules only have two sensors their average noise of about 
 1000~\eenc is lower than the noise for middle and outer modules, 
 which is similar and on average amounts to 1350~\eenc and 1450\eenc.
 The temperature dependence is about 6~\eenc per degree Celsius,
 the gain is about 50~\mvfc.
 Since the average signal is about 3.3~\fc, the signal to noise ratio, 
 e.g.~for middle modules, is about 15 for non-irradiated modules.
 \par
\begin{figure}[htb]
\begin{center}
{\includegraphics[width=0.49\linewidth,clip]{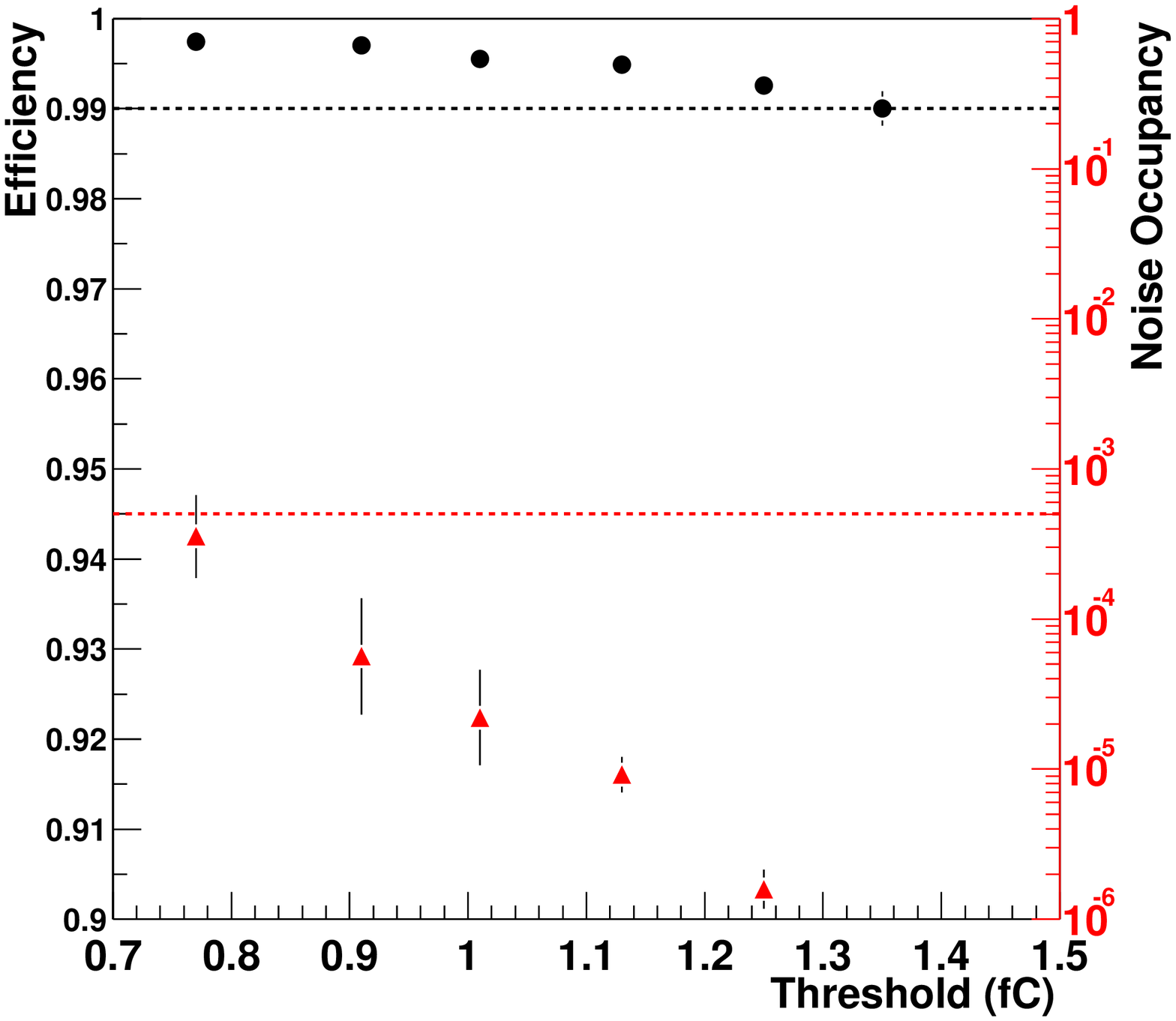}}
{\includegraphics[width=0.49\linewidth,clip]{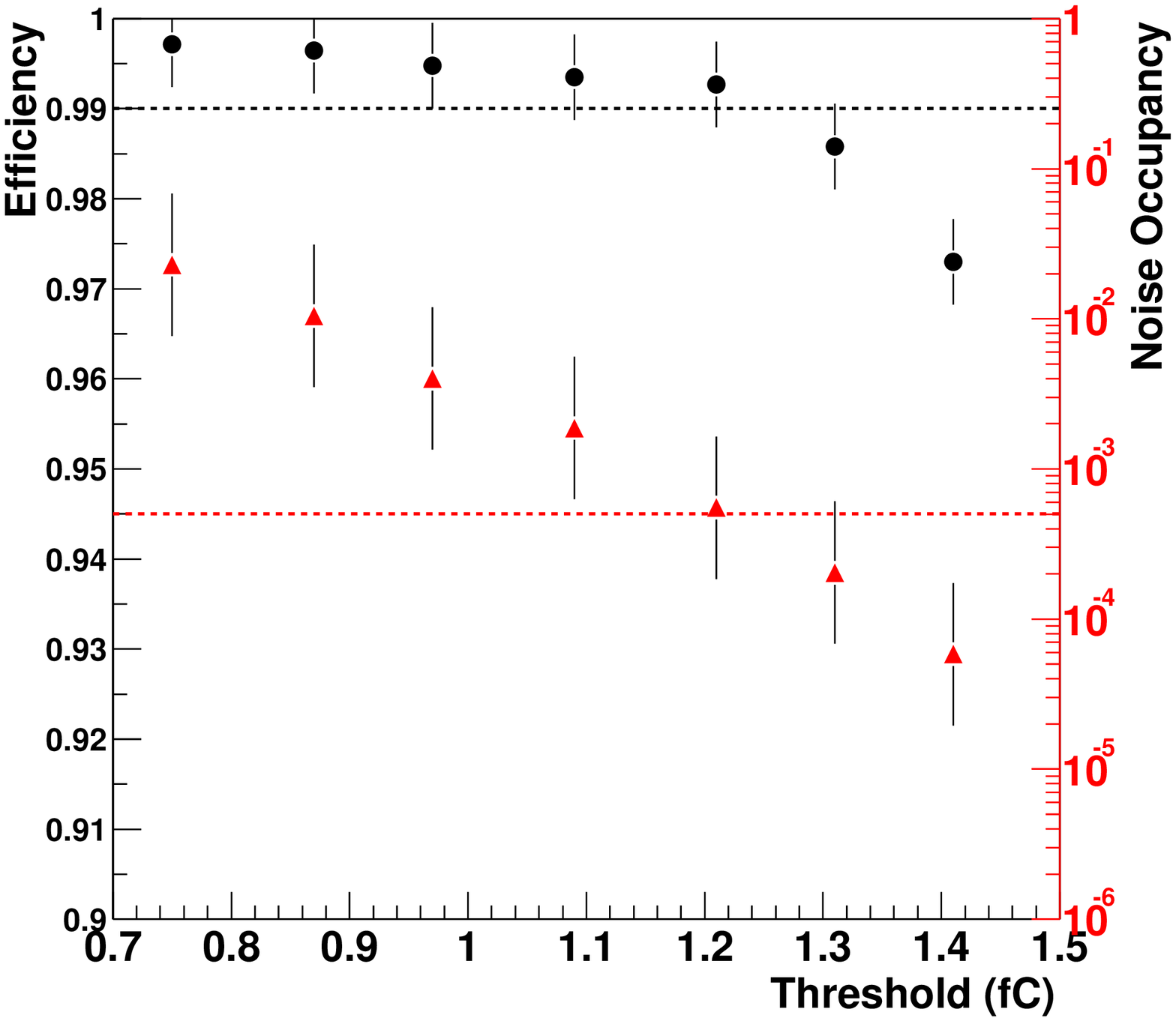}}
\begin{picture}(0,0)
\put(-355,30){\large\bf a)}
\put(-160,30){\large\bf b)}
\end{picture}
\caption{{\it The efficiency and noise occupancy versus discriminator 
              threshold for a) a non-irradiated and b) a fully irradiated  
              prototype end-cap module.}
        }\label{fig04}
\end{center}
\end{figure}

 The best performance is reached at as high as possible efficiency with 
 as low as possible noise occupancy.
 For non-irradiated modules, these two quantities are shown
 in Fig.~\ref{fig04}a  as functions of the discriminator threshold.
 The design values of an efficiency larger than 99$\%$ at a noise occupancy
 of less than $5 \cdot 10^{-4}$ can be reached for a large range of 
 operating thresholds.
 However, the module performance is deteriorated after receiving the 
 full dose of ten years of operation at the LHC.
 The radiation damage at the LHC is conservatively simulated by 
 irradiating the modules with 24~\gev protons with a fluence of 
 $3.3\cdot10^{14}$ p/cm$^2$.
 After this treatment the module performance has been significantly 
 degraded~\cite{ELEC}.
 The leakage current rises from a few nA to about 1.5~mA at 400~V for outer
 modules, and now significantly contributes to the heat budget of the modules.
 The noise is increased, e.g.~for outer modules to about 2100-2400~\eenc,
 and the temperature dependence raises to about 20~\eenc per degree Celsius.
 The collected charge decreases, and the gain drops to about 30~\mvfc.
 The radiation damage results in a lower efficiency at larger noise 
 occupancy and the design values can only be reached in a very small 
 window of operating threshold around 1.2~\fc, Fig.~\ref{fig04}b.
 \par
%
%------------------------------------------------------------------------------
\begin{figure}[htb]
\begin{center}
{\includegraphics[width=1.00\linewidth,clip]{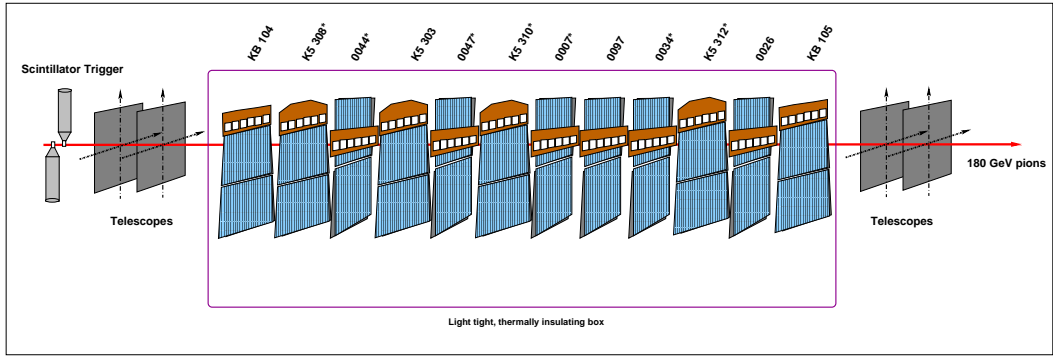}}
\caption{{\it Schematics of the test beam setup. Fully irradiated modules
              are marked with a star.}
        }\label{fig05}
\end{center}
\end{figure}

%------------------------------------------------------------------------------
%
 To verify that, even with the degraded performance expected
 towards the end of LHC running, high tracking
 efficiencies at low fake rates can still be obtained, several fully 
 irradiated modules, interspersed with non-irradiated ones, are put 
 into a test beam setup, Fig.~\ref{fig05}.
 \par
%
%------------------------------------------------------------------------------
%------------------------------------------------------------------------------
%
 The fully irradiated barrel and end-cap 
 modules, marked with a star in Fig.~\ref{fig05}, 
 are placed at about the distances they will have in ATLAS.
 The position of the 180~\gev pion beam is precisely determined using
 an external silicon telescope with analog read-out. 
 The tracking performance is studied by varying the noise occupancy via 
 the threshold~\cite{BEAM}.
 The observed residuals of the space points per module are according to 
 expectations from the geometry, about 17 (800)~$\mu m$ perpendicular
 (parallel) to the strips.
 The tracking efficiency is found to be similar for barrel and 
 end-cap modules, and to be consistent with the product of the 
 individual sensor efficiencies.
 For a track reconstructed from three modules, and at 1.2~\fc threshold, an 
 efficiency of more than 97$\%$ at about 10$^{-3}$ fake rate is achieved.
 For four modules the efficiency is still larger than 97$\%$, however at 
 a lower fake rate compatible with zero.
 In conclusion, this demonstrates that the fully irradiated modules still 
 allow for tracking with high efficiency and low fake rate.
%
%------------------------------------------------------------------------------
%
\section{SCT production status}
\label{sec:produ}
 The SCT Collaboration consists of 284 members from 41 institutions in 
 Asia, Australia, Europe and the US.
 Consequently, both for the barrel and the end-cap part, the production of 
 modules is distributed among several production centers around the world,
 requiring a complex logistics. The control of material and test results 
 is provided by a central database~\cite{DBASE}.
 Mounting of modules to barrels and end-cap disks is done in Japan,
 the UK and the Netherlands.
 The final integration into the ATLAS inner detector will be performed 
 at CERN.
 For the barrel part module production is well underway and about 
 1/3 completed, whereas the end-cap module production has just started.
 The installation of services on barrels and disks has commenced, 
 but no module mounting has been done as of now.
 The envisaged completion dates are end of 2004 for the barrel and mid 
 2005 for the end-caps.
%
%------------------------------------------------------------------------------
%
\section{Conclusions}
\label{sec:concl}
 The ATLAS inner detector will be equipped with a silicon strip
 detector, the SCT.
 A number of prototype end-cap modules demonstrates that the mechanical 
 requirements can be met with sufficient yield.
 The electrical performance of non-irradiated modules is according
 to the design.
 In contrast, the irradiated modules only allow for a marginal
 operating flexibility after receiving the full LHC dose.
 The series production is underway for barrel modules and has just
 started for end-cap modules.
 In December 2004 (May 2005) the barrel (end-cap) part is expected to be ready
 for integration into the ATLAS inner detector.
 \\\mbox{ }\\
%
%------------------------------------------------------------------------------
%
 \appendix
 {\bf Acknowledgement:}\\
 I wish to thank the organisers of this interesting conference
 for the fruitful atmosphere they created throughout the meeting.
%
%------------------------------------------------------------------------------
%

\end{document}